\documentclass[twocolumn,pre,showpacs,preprintnumbers,eqsecnum,amssymb]{revtex4}

\usepackage{graphicx}
\usepackage{dcolumn}
\usepackage{bm}

\begin{document}

\preprint{asymmetric-br.tex}

\title{Density profiles of Ar adsorbed in slits of CO$_2$:
spontaneous symmetry breaking}

\author{Leszek Szybisz}

\affiliation{Laboratorio TANDAR, Departamento de F\'{\i}sica,
Comisi\'on Nacional de Energ\'{\i}a At\'omica,\\
Av. del Libertador 8250, RA--1429 Buenos Aires, Argentina}
\affiliation{Departamento de F\'{\i}sica, Facultad de
Ciencias Exactas y Naturales,\\
Universidad de Buenos Aires,
Ciudad Universitaria, RA--1428 Buenos Aires, Argentina}
\affiliation{Consejo Nacional de Investigaciones
Cient\'{\i}ficas y T\'ecnicas,\\
Av. Rivadavia 1917, RA--1033 Buenos Aires, Argentina}

\author{Salvador A. Sartarelli}

\affiliation{Instituto de Desarrollo Humano, Universidad
Nacional de General Sarmiento, Gutierrez 1150,
RA--1663 San Miguel, Argentina}

\date{\today}

\begin{abstract}
A recently reported symmetry breaking of density profiles of fluid
argon confined by two parallel solid walls of carbon dioxide is
studied. The calculations are performed in the framework of a
nonlocal density functional theory. It is shown that the existence
of such asymmetrical solutions is restricted to a special choice
for the adsorption potential, where the attraction of the solid-fluid
interaction is reduced by the introduction of a hard-wall repulsion.
The behavior as a function of the slit's width is also discussed.
All the results are placed in the context of the current knowledge
on this matter.

\end{abstract}


\maketitle

\section{Introduction}
\label{sec:introduce}

In a quite recent paper, Berim and Ruckenstein \cite{berim07}
have reported symmetry breaking of the density profile of fluid
argon (Ar) confined in a planar slit with identical walls of
carbon dioxide (CO$_2$). These authors claimed that a completely
symmetric integral equation provides an asymmetric profile which
has a lower free energy than that of the lowest symmetric solution
leading to a symmetry breaking phenomenon. It was assumed that the
Ar atoms interact via a standard Lennard-Jones (LJ) potential
characterized by the strength $\varepsilon_{ff}$ and the atomic
diameter $\sigma_{ff}$. The presented results were obtained from
calculations carried out with the smoothed density approximation
(SDA) version \cite{taraz85,taraz87} of the nonlocal density
functional theory (DFT) in the case of a closed planar slit with an
effective width of $15\,\sigma_{ff}$. The symmetry breaking was
found for temperatures between the experimental triple point for Ar,
$T_t=83.8$ K, and a critical value $T_{sb}=106$ K. At each
temperature, it was determined a range of average densities
$\rho_{sb1} \le \rho_{av} \le \rho_{sb2}$ where the symmetry breaking
occurs, outside this range a symmetric profile has the lowest free
energy.

As a matter of fact, the adsorption of fluid Ar on a solid substrate
of CO$_2$ was intensively studied for several decades (see, e.g.,
Refs.\ \onlinecite{nicholson} and \onlinecite{brush}). In 1977 Ebner
and Saam \cite{ebner77} analyzed phase transitions by assuming that
atoms of the fluid interact with the solid wall via a 9-3 van der
Walls potential (from here on denoted as ES potential) obtained
from the assumption that Ar atoms interact with CO$_2$ molecules
via a LJ interaction with parameters $\varepsilon_{sf}$ and
$\sigma_{sf}$. After this pioneering work, a large amount of work
has been devoted to study this system with different numerical and
analytical techniques. The attention was focused to analyze
features like: the oscillatory behavior of the density profile
which leads to a layered structure in the neighborhood of the flat
substrate; the thin- to thick-film transitions; wetting properties
and prewetting jumps \cite{sullivan79,johnson80,foiles82,taraz82,%
evans83,meister85,bruno87,finn89,sokolow90,velasco90,finn90,fan93a,%
fan93b,kierlik94,kierlik95,mistura99}. Berim and Ruckenstein 
\cite{berim07} have also adopted the ES potential, however, a
hard-wall repulsion was introduced in their calculations. In
practice, such a hard wall diminishes the strength of the
solid-fluid interaction. 

The investigation of symmetry breaking in physical systems is a
very exciting issue. This is due to the fact that such a feature
may have fundamental theoretical implications. In many fields of
physics the discovery of symmetry breaking lead to significant
advances in the theory. Therefore it is important to place the
results of Ref.\ \cite{berim07} in the context of the current
knowledge about adsorption into planar slits.

Asymmetric solutions for fluids confined in slits have been
previously reported in the literature. About 20 years ago a Dutch
Collaboration has carried out calculations on the Delft Molecular
Dynamics Processor (DMDP), which was specially designed for
Molecular Dynamics (MD) simulations of simple fluids
\cite{bakker83,bakker88}. The results were published in a series
of papers by Sikkenk et al. \cite{sikkenk87,sikkenk88} and
Nijmeijer et al. \cite{nijmeijer89,nijmeijer90}. The simulations
were performed for a canonical ensemble with two types of
particles, 2904 of one type for building a solid substrate and
several thousand of the other type for composing the fluid adsorbate.
The temperature of the system was kept at $T^* = k_B\,T/
\varepsilon_{ff}=0.9$ which is in between the fluid's triple-point
temperature $T^*_t \simeq 0.7$ and the critical temperature
$T^*_c \simeq 1.3$. The width of the slit was taken as $L=29.1\,
\sigma_{ff}$, supposing that this distance be enough large to avoid
any capillary effect. Such a system can support solid-liquid (SL),
solid-vapor(SV), and liquid-vapor (LV) interfaces. These authors
have studied wetting at LV coexistence by varying over a wide range
the relative strength of the solid-fluid and fluid-fluid
interactions defined by the ratio $\varepsilon_{r} =
\varepsilon_{sf}/\varepsilon_{ff}$ of the LJ parameters. The length
scale of this interaction was taken as $\sigma_{sf}=0.941\,
\sigma_{ff}$. For increasing $\varepsilon_{r}$ from $\varepsilon_{r}
\simeq 0.1$ towards $\varepsilon_{r} \simeq 1.0$ three cases were
observed:
 
(i) at low $\varepsilon_{r}$, {\it symmetric} profiles consisting of
two SV interfaces and two LV interfaces are obtained, this situation
corresponds to a complete wall drying as can be seen in Fig.\ 1 of
\cite{nijmeijer90};

(ii) at intermediate $\varepsilon_{r}$, {\it asymmetric} profiles
consisting of a SL, a LV, and a SV interface are obtained, here the
wall attraction is sufficiently strong to produce a partial wetting,
i.e., to support a rather thick film on one wall while a SV interface
is present near the other wall, this feature is shown in Fig.\ 2 of
\cite{nijmeijer90};

(iii) for the largest $\varepsilon_{r}$, {\it symmetric} profiles
consisting of two SL interfaces and two LV interfaces are obtained,
now the strength is enough to wet both walls as can be seen in Fig.\
3 of \cite{nijmeijer90}.

The structure of the profiles mentioned above depends on the balance
of the involved surface tensions $\gamma_{SL}$, $\gamma_{LV}$, and
$\gamma_{SV}$ which are related by the Young's law (see, e.g., Eq.\
(2.1) in Ref.\ \cite{deGennes85})
\begin{equation}
\gamma_{SV} = \gamma_{SL} +\gamma_{LV}\,\cos{\theta}  \:,
\label{young}
\end{equation}
where $\theta$ is the contact angle. The latter quantity is defined
as the angle between the wall and the interface between the liquid
and the vapor (see Fig. 1 in Ref.\ \cite{deGennes85}). The
transition from (i) to (ii) takes place at the drying point $\theta
=\pi$, whereas the transition from (ii) to (iii) takes place at the
wetting point $\theta=0$.  

It is worth of notice that Velasco and Tarazona \cite{velasco89}
have carried out calculations in the framework of the SDA obtaining
density profiles with the same structure to that reported in Refs.\
\cite{sikkenk87,sikkenk88,nijmeijer89,nijmeijer90}. The reader may
look at Ref.\ \cite{henderson91} for a further comparison between
MD and DFT results.

It is the aim of the present work to acquire a more accurate picture
of the symmetry breaking reported in Ref. \onlinecite{berim07}. In so
doing, we explore size effects by comparing the results obtained for
slits of widths $15\,\sigma_{ff}$ and $30\,\sigma_{ff}$. Next, we
investigate the existence of stable asymmetric solutions for the
density profiles when the position of the hard-wall repulsion
introduced in Ref.\ \onlinecite{berim07} is changed. When the
location of this hard wall is moved, the strength of the adsorption
potential is varied allowing a connection to the studies described
in Refs.\ \cite{sikkenk87,sikkenk88,nijmeijer89,nijmeijer90,%
velasco89,henderson91}. Several properties of the obtained solutions
are discussed.

The paper is organized as follows. In Sec. \ref{sec:model} we
provide a summary of the formalism underlying the present
calculations. Special attention is devoted to the adsorption
potential. The results and its analysis are reported in Sec.
\ref{sec:results}. Final remarks are given in Sec.\
\ref{sec:conclude}.

\section{The model}
\label{sec:model}

The properties of a fluid adsorbed by an inert solid substrate may
be studied by analyzing the grand free energy \cite{ravikovitch01}
\begin{equation}
\Omega = F - \mu\,N \:, \label{omega00}
\end{equation}
where $F$ is the Helmholtz free energy, $\mu$ the chemical
potential, and $N$ the number of particles of the adsorbate
\begin{equation}
N = \int \rho({\bf r})\,d{\bf r} \;. \label{particle}
\end{equation}
Quantity $F$ contains the energy due to the interaction between
fluid atoms as well as the energy provided by the confining
potential. In a DFT it is expressed in terms of the density
profile $\rho({\bf r})$
\begin{equation}
F[\rho({\bf r})] = F_{\rm int}[\rho({\bf r})]
+ \int d{\bf r}\,\rho({\bf r})\,U_{sf}({\bf r})
\;. \label{free0}
\end{equation}
Here $F_{\rm int}[\rho(\bf r)]$ is the intrinsic Helmholtz free
energy functional and $U_{sf}({\bf r})$ is the external potential
produced by the slit's walls.

This formulation is usually applied to systems described by the
grand canonical ensemble, i.e., at constant volume $V$, temperature
$T$, and chemical potential $\mu$. Such a situation corresponds to
an open system in contact with reservoir which fixes $T$ and $\mu$.
A minimization of $\Omega$ with respect to $\rho({\bf r})$ leads
to the Euler-Lagrange equation for the density profile and the
number of particles may be evaluated with Eq.\ (\ref{particle}).
For a closed system, i.e., a canonical ensemble with fixed $N$,
one should treat $\mu$ as an unknown Lagrange multiplier to be
determined from the minimization procedure.

\subsection{Density functional theory}
\label{sec:functional}

Let us now summarize the DFT adopted for $F_{\rm int}[\rho({\bf
r})]$. In the case of inhomogeneous classical fluids at temperature
$T$ the intrinsic free energy functional is decomposed into two
kind of contributions:

(i) the ideal gas term $F_{\rm id}[\rho({\bf r})]$, which is given
by the exact expression
\begin{eqnarray}
F_{\rm id}[\rho({\bf r})] &=& k_B\,T \int d{\bf r}\,\rho({\bf r})
\,f_{\rm id}[(\rho)] \nonumber\\
&=& k_B\,T \int d{\bf r}\,\rho({\bf r})\,
\{\ln[\Lambda^3\rho({\bf r})]-1\} \;, \label{free1}
\end{eqnarray}
with $\Lambda=\sqrt{2\,\pi\,\hbar^2/m\,k_B\,T}$ being the thermal
de Broglie wavelength of a molecule of mass $m$;

(ii) the excess term $F_{\rm ex}[\rho({\bf r})]$, which accounts for the interparticle interactions is a unique but unknown
functional of the local density. For fluids with attractive
interactions as the Lennard-Jones (LJ) one, the free energy is
decomposed into the repulsive and attractive contributions. The
repulsive interactions are then approximated by a hard-sphere
functional with a certain choice of the hard-sphere diameter
$d_{\rm HS}$
\begin{equation}
F_{\rm HS}[\rho({\bf r})] = \int d{\bf r}\,\rho({\bf r})
\,f_{\rm HS}[\bar\rho({\bf r});d_{\rm HS}]
\;, \label{hard0}
\end{equation}
whereas the attractive interactions are treated in most cases in
a mean field fashion
\begin{equation}
F_{\rm attr}[\rho({\bf r})] = \frac{1}{2} \int \int
d{\bf r}\,d{\bf r\prime}\,\rho({\bf r})\,\rho({\bf r\prime})\,
\Phi_{\rm attr}(\mid {\bf r} - {\bf r\prime} \mid)
\;. \label{free4}
\end{equation}
Here $\Phi_{\rm attr}(r=\mid {\bf r} - {\bf r\prime} \mid)$ is the
attractive part of the LJ potential.

In summary, the intrinsic Helmholtz free energy functional may be
expressed as
\begin{eqnarray}
F_{\rm int}[\rho({\bf r})]
&=& F_{\rm id}[\rho({\bf r})] + F_{\rm ex}[\rho({\bf r})]
\nonumber\\
&=& F_{\rm id}[\rho({\bf r})] + F_{\rm HS}[\rho({\bf r})]
+ F_{\rm attr}[\rho({\bf r})] \;. \label{free2}
\end{eqnarray}
The free energy functional for hard spheres plays a central role
in DFT. Expressions for $f_{\rm HS}[\bar\rho({\bf r});d_{\rm HS}]$
may be taken from the Percus-Yevick \cite{percus58} or
Carnahan-Starling (CS) \cite{carnahan69} approximations for the
equation of state (EOS) of a uniform non-attractive hard-sphere
fluid (see, e.g. Ref.\ \cite{barker76}). In a nonlocal DFT this
quantity is evaluated as a function of a conveniently averaged
density $\bar\rho({\bf r})$.

For the calculations performed in the present work we used the
same SDA formalism adopted in the paper of Berim and Ruckenstein
\cite{berim07}. In this approach developed by Tarazona
\cite{taraz85,taraz87}, the excess of free energy density of hard
spheres is written according to the semi-empirical quasi-exact
CS expression
\begin{equation}
f_{\rm HS}[\bar\rho({\bf r});d_{\rm HS}]] = f_{\rm CS}[\bar\eta]
= k_B\,T\,\biggr[\frac{\bar\eta\,(4-3\bar\eta)}{(1-\bar\eta)^2}
\biggr] \;. \label{free6}
\end{equation}
Here $\bar\eta=\bar\rho({\bf r})\,V_{\rm HS}$ is the packing
fraction, where the factor $V_{\rm HS}=\pi\,d^3_{\rm HS}/6$ is
the volume of a hard sphere. The smoothed density $\bar\rho({\bf
r})$ is defined as
\begin{equation}
\bar\rho({\bf r}) = \int d{\bf r}\,\rho({\bf r})
\,w[\mid {\bf r} - {\bf r\prime} \mid ; \bar\rho({\bf r})]
\;, \label{hard1}
\end{equation}
with the following weighting function:
\begin{eqnarray}
w[ \mid {\bf r} - {\bf r\prime} \mid ; \bar\rho({\bf r})]
&=& w_0[ \mid {\bf r} - {\bf r\prime} \mid ]
+ w_1[ \mid {\bf r} - {\bf r\prime} \mid ]\,\bar\rho({\bf r})
\nonumber\\
&&+ w_2[ \mid {\bf r} - {\bf r\prime} \mid ]
\,\bar\rho^2({\bf r}) \;. \label{weight}
\end{eqnarray}
The expansion coefficients $w_0(r)$, $w_1(r)$, and $w_2(r)$ are
density independent and its expressions as a function of
$r = \mid {\bf r} - {\bf r\prime} \mid$ are given in the
Appendix of Ref.\ \onlinecite{taraz87}. 
 
To account for the fluid-fluid interaction we adopted, as in Ref.\ 
\cite{berim07}, the spherically symmetric L-J potential given
in Eq. (2) of Ref.\ \onlinecite{johnson80}
\begin{eqnarray}
\Phi_{\rm attr}(r)
 = \left\{ \begin{array}{ll}
   4 \varepsilon_{ff} \biggr[ \left(\frac{\sigma_{ff}}{r}
   \right)^{12} - \left(\frac{\sigma_{ff}}{r}\right)^6 \biggr]
      & \mbox{if}\;\;r \ge \sigma_{ff} \, \\
   0  & \mbox{if}\;\;r < \sigma_{ff}
        \end{array} \right. \;. \label{jones}
\end{eqnarray}
where $\sigma_{ff}$ is the hard-core diameter of the fluid. The
authors of Ref.\ \onlinecite{johnson80} have used this L-J version
just for studying the adsorption of Ar on CO$_2$ and it has been also
utilized in several subsequent works on this system. The values of
the interaction parameters for Ar are $\varepsilon_{ff}/k_B =
119.76$~K and $\sigma_{ff}=3.405$ \AA.

\subsection{The Euler-Lagrange equation}
\label{sec:equation}

The equilibrium density profile $\rho({\bf r})$ of the fluid
adsorbed in a closed slit is determined by a minimization
of the free energy with respect to density variations with the
constraint of a fixed number of particles $N$
\begin{eqnarray}
\frac{\delta}{\delta \rho({\bf r})} \biggr[
F_{\rm int}[\rho({\bf r})] + \int d{\bf r\prime}\,
\rho({\bf r\prime})\,[U_{sf}({\bf r\prime}) - \mu] \biggr] = 0
\;. \nonumber \\ \label{euler-l}
\end{eqnarray}
Here, i.e., for an ensemble with fixed $V$, $T$, and $N$, the
Lagrange multiplier $\mu$ is an unknown quantity which should be
determined from the constraint. It plays a role of a chemical
potential but off the liquid-vapor coexistence conditions. Hence,
it is not necessarily equal to $\mu_{coex}$ of an open slit in
equilibrium with a reservoir at temperature $T$ (see, e.g., Ref.\
\onlinecite{ancilotto99}).

In the case of a planar symmetry where the flat walls exhibit an
infinite extent in the $x$ and $y$ directions the profile depends
only of the coordinate $z$ perpendicular to the substrate. For
this geometry the variation of Eq.\ (\ref{euler-l}) yields the
following Euler-Lagrange (EL) equation
\begin{eqnarray}
\frac{\delta F}{\delta \rho(z)} &+& U_{sf}(z) 
= \frac{\delta [F_{\rm id} + F_{\rm HS}]}{\delta \rho(z)}
\nonumber\\
&+& \int dz\prime\,\rho(z\prime)\,
\bar\Phi_{\rm attr}(\mid z - z\prime \mid) + U_{sf}(z) = \mu \;.
\nonumber\\ \label{euler}
\end{eqnarray}
For a slit of effective width $\ell_w$ this EL equation may be cast
into the form
\begin{eqnarray}
k_B\,T\,\ln{[\Lambda^3\,\rho(z)]} + Q(z) = \mu \;, \label{euler1}
\end{eqnarray}
where
\begin{eqnarray}
Q(z)&=&k_B\,T\,
\frac{4\,\bar\eta(z)-3\bar\eta^2(z)}{[1-\bar\eta(z)]^2}
\nonumber\\
&&+ k_B\,T\,\frac{\pi\,d^3_{HS}}{6}\,
\int^{\ell_w}_0 dz\prime\,\rho(z\prime)\,
\frac{4-2\,\bar\eta(z\prime)}{[1-\bar\eta(z\prime)]^3}
\,\frac{\delta \bar\rho(z\prime)}{\delta \rho(z)}
\nonumber\\
&&+ \int^{\ell_w}_0 dz\prime\,\rho(z\prime)\,
\bar\Phi_{\rm attr}(\mid z - z\prime \mid) + U_{sf}(z)
\;. \label{euler2}
\end{eqnarray}
The number of particles per unit area of one wall of the slit is
\begin{equation}
N_s = \int^{\ell_w}_0 \rho(z)\,dz \;. \label{cover}
\end{equation}

In order to get solutions for $\rho(z)$ it is useful to rewrite
Eq.\ (\ref{euler1}) as
\begin{equation}
\rho(z) = \rho_0\,\exp{\left(-\frac{Q(z)}{k_B\,T}\right)} \;,
\label{rho}
\end{equation}
with
\begin{equation}
\rho_0 = \frac{1}{\Lambda^3}\,
\exp{\left(\frac{\mu}{k_B\,T}\right)} \;. \label{rho0}
\end{equation}
The relation between $\mu$ and $N_s$ is obtained by substituting
Eq.\ (\ref{rho}) into the constraint given by Eq.\ (\ref{cover})
\begin{equation}
\mu = - k_B\,T\,\ln{\biggr[\frac{1}{N_s\,\Lambda^3}
\int^{\ell_w}_0 dz\,\exp{\left(-\frac{Q(z)}{k_B\,T}\right)
\biggr]}} \;. \label{lambda}
\end{equation}
For the calculations carried out in the present work we set
$d_{\rm HS} = \sigma_{ff}$ as it was done in Ref.\
\onlinecite{berim07}.

The asymmetry of the density profiles is measured by the parameter
\begin{equation}
\Delta_N = \frac{1}{2\,N_s} \int^{\ell_w}_0 dz \mid \rho(z)
- \rho(\ell_w-z) \mid \;. \label{asymm}
\end{equation}
According to this definition, if the profile is completely
asymmetrical about the middle of the slit [$\rho(z <\ell_w/2) \ne 0$
and $\rho(z \ge \ell_w/2) = 0$] this parameter becomes unity, while
for symmetric solutions it vanishes.

\subsection{Adsorption potential}
\label{sec:potential}

The model van der Waals (9-3) potential proposed by Ebner and
Saam \cite{ebner77}, i.e. the ES potential, is
\begin{eqnarray}
U_{sf}(z)
 = \left\{ \begin{array}{ll}
   \frac{2 \pi}{3} \epsilon_{eff}
   \biggr[\frac{2}{15}\left(\frac{\sigma_{sf}}{z}\right)^9
   - \left(\frac{\sigma_{sf}}{z}\right)^3 \biggr]
      & \mbox{if}\;\;z > 0 \, \\
   0  & \mbox{if}\;\;z \le 0
        \end{array} \right. \, \label{plane0}
\end{eqnarray}
with $\epsilon_{eff}=\varepsilon_{sf}\,\rho_{s}\,\sigma^3_{sf}$
being the effective strength, was adopted for almost all the
abovementioned studies of the adsorption of Ar atoms on a flat
wall of solid CO$_2$. The exception is the experimental and
theoretical investigation performed by Mistura {\it et al.}
\cite{mistura99}, where a more realistic adsorption potential
calculated on the basis an {\it ab initio} expansion of Marshall
{\it et al.} \cite{marshall96} was used. The ES expression is
obtained when one assumes that Ar atoms interact with CO$_2$ atoms
via a Lennard-Jones (12-6) potential and subsequently integrates
this potential over a continuum of CO$_2$ substrate atoms with a
reduced density $\rho^*_s=\rho_{s}\,\sigma^3_{sf}=0.988$. The
cross-parameters of the potential are determined by using the
Lorentz-Berthelot rules. So that, the van der Waals strength
$\varepsilon_{sf}$ is the square root of the product of the
argon and CO$_2$ van der Waals strengths, while the hard-core
diameter $\sigma_{sf}$ is the mean of the argon and CO$_2$
hard-core diameters, while The parameters evaluated in this way
are $\varepsilon_{sf}/k_B=153$~K and $\sigma_{sf}=3.727$ \AA.

\begin{figure}
\centering\includegraphics[width=8cm, angle=-0]{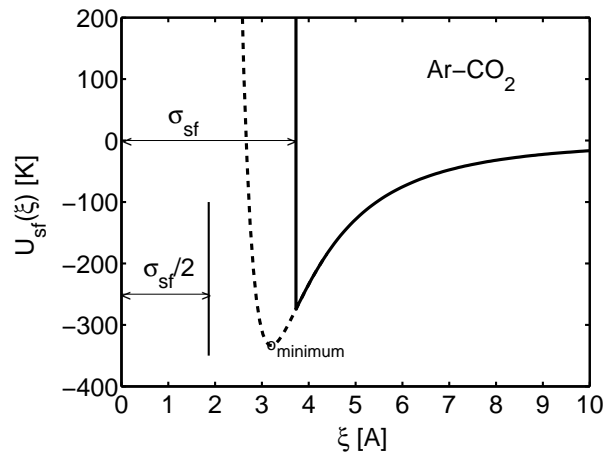}
\caption{\label{fig:potential} Adsorption potential as a function
of the distance from the real wall. The solid curve is the
potential of Berim and Ruckenstein close to the left wall given
by Eq.\ (\ref{sberim}), while the dashed curve is that of 
Nilson and Griffiths given by Eq.\ (\ref{snilson}).}
\end{figure}

Berim and Ruckenstein \cite{berim07} have investigated the
Ar-CO$_2$ system utilizing, in principle, the ES potential.
However, by looking at their paper one realizes that according
to Eq. (A5) of the Appendix
\begin{equation}
U_{sf1}(z) = \frac{2 \pi}{3} \epsilon_{eff} \biggr[
   \frac{2}{15}\left(\frac{\sigma_{sf}}{z+\sigma_{sf}}\right)^9
 - \left(\frac{\sigma_{sf}}{z+\sigma_{sf}}\right)^3
\biggr] \;, \label{pberim}
\end{equation}
which accounts for the solid-fluid interaction at one of the
walls, a hard-wall repulsion was located at a distance
$\sigma_{sf}$ from the real wall of the slit. In agreement with
this assumption, the total confining potential exerted on Ar
atoms by the two walls separated by a distance $L$ was expressed
as
\begin{equation}
U_{sf}(z) = U_{sf1}(z+\sigma_{sf}) + U_{sf2}(\ell_w-z+\sigma_{sf})
\;. \label{sberim}
\end{equation}
Here the effective width of the slit is
\begin{equation}
\ell_w = L - 2\,\sigma_{sf} \;. \label{lberim}
\end{equation}
This scenario is depicted in Fig. 2 of Ref.\ \onlinecite{berim07}.
In this context, it is interesting to notice that Nilson and
Griffiths \cite{nilson99} in order to study the adsorption of a
fluid in a planar slit have written in their Eq.\ (10) the total
fluid-solid potential as
\begin{equation}
U_{sf}(z) = U_{sf1}(z+\frac{\sigma_{sf}}{2})
+ U_{sf2}(\ell_w-z+\frac{\sigma_{sf}}{2}) \;, \label{snilson}
\end{equation}
i.e., locating a hard-wall repulsion at a distance $\sigma_{sf}/2$
from the substrate. In this case, the effective width is
\begin{equation}
\ell_w = L - \sigma_{sf} \;, \label{lnilson}
\end{equation}
as it is shown in Fig. 2 of Ref.\ \onlinecite{nilson99}.

In Fig.\ \ref{fig:potential} we compare the potentials outlined in
the previous paragraph. The comparison is restricted to the region
close to the substrate. The quantity $\xi$ is the perpendicular
distance from the real wall being
\begin{eqnarray}
\xi
 = \left\{ \begin{array}{ll}
   z+\sigma_{sf} & \mbox{for}\;\; {\rm Berim-Ruckenstein} \, \\
   \, \\
   z+\frac{\sigma_{sf}}{2} & \mbox{for}\;\; {\rm Nilson-Griffiths.}
        \end{array} \right. \label{ksi}
\end{eqnarray}
One may realize that Eq.\ (\ref{snilson}) retains the ``soft''
repulsion [$U_{sf}(z) \propto (\sigma_{sf}/z)^9$], while Eq.\
(\ref{sberim}) cuts the potential before the minimum be reached.
This feature produces important effects on the behavior of the
density profiles. In fact, the calculations performed by Berim and
Ruckenstein \cite{berim07} yielded density profiles with $\rho(z=0)$
and $\rho(z=\ell_w)$ different from zero indicating that the fluid
is in contact with the hard walls, while in the case of Nilson and
Griffiths \cite{nilson99} the fluid forms a well defined first
layer separated from the wall.
 
In the present work we shall analyze the evolution of asymmetric
solutions when the total adsorption potential is written as
\begin{equation}
U_{sf}(z) = U_{sf1}(z+\frac{\sigma_{sf}}{\nu})
+ U_{sf2}(\ell_w-z+\frac{\sigma_{sf}}{\nu}) \;, \label{pown}
\end{equation}
and the parameter $\nu$ varies from 1 to 2. In doing so, one goes
from Eq.\ (\ref{sberim}) towards Eq.\ (\ref{snilson}) increasing the
strength of the solid-fluid attraction.

\begin{table}
\caption{\label{table0} Values of the Helmholtz free energies
$F_{sym}$, $F_{asym}$, and $F_{cap}$ of the symmetric, asymmetric
and capillary solutions, respectively, obtained with $\nu=1$ for the
slit $\ell^*_w=15$ at $T=87$ K. The free energies are given for
several average densities in the range $\rho^*_{sb1} \le
\rho^*_{av} \le \rho^*_{sb2}$ in units of $k_B\,T/\sigma^2_{ff}$.}
\begin{ruledtabular}
\begin{tabular}{cccccc}
&\multicolumn{2}{c}{$F_{sym}$}&\multicolumn{2}{c}{$F_{asym}$}
& $F_{cap}$ \\
$\rho^*_{av}$ & BR$^a$ & PW$^b$ & BR$^a$ & PW$^b$ & PW$^b$ \\
\hline
\\
0.1546 & $-26.59$ & $-26.62$ & $-26.67$ & $-26.70$ & \\
0.2319 & $-39.66$ & $-39.69$ & $-39.86$ & $-39.88$ & $-39.12$ \\
0.3092 & $-52.81$ & $-52.85$ & $-53.06$ & $-53.08$ & $-52.32$ \\
0.3865 & $-66.01$ & $-66.04$ & $-66.26$ & $-66.29$ & $-65.65$ \\
0.4638 & $-79.22$ & $-79.28$ & $-79.48$ & $-79.51$ & $-79.26$ \\
\end{tabular}
\end{ruledtabular}
$^a$ Data taken from \cite{berim07}. \\
$^b$ Calculated in the present work. \\
\end{table}

\begin{figure}
\centering\includegraphics[width=8cm, height=6cm]{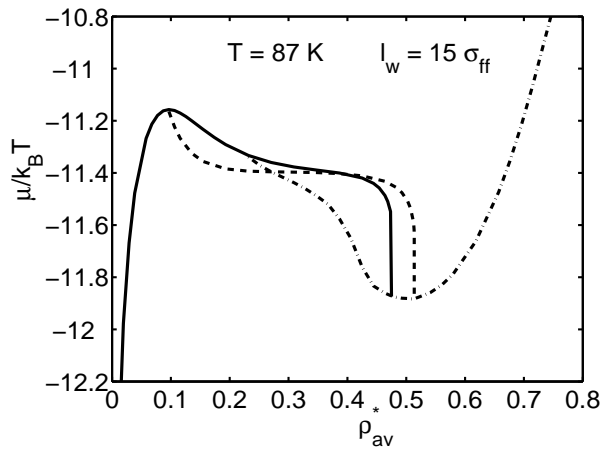}
\caption{\label{fig:chemical} Lagrange multiplier $\mu$ as a
function of average density. The solid curve are results for
symmetric film solutions. The dashed curve stands for values of
asymmetric film solutions which occur in the range $\rho^*_{sb1}
\le \rho^*_{av} \le \rho^*_{sb2}$. The dot-dashed curve corresponds
to drying-CC like solutions.}
\end{figure}

\section{Numerical results and discussion}
\label{sec:results}

Let us now describe the obtained results. The EL equation
(\ref{euler1}) was solved at fixed $\ell_w$ and $T$ for a given
number of particles per unit area $N_s$. The latter quantity
determines an average fluid density $\rho_{av}=N_s/\ell_w$. A widely
used computational algorithm consisting of a numerical iteration of
the coupled Eqs.\ (\ref{rho})-(\ref{lambda}) was applied. This
procedure yields the density profile $\rho(z)$ and the value of the
Lagrange multiplier $\mu$. The convergence of the solutions are
measured by the difference between two consecutive profiles
\begin{equation}
\delta_1 = \sigma^5_{ff}\, \int^{\ell_w}_0 dz \biggr[
\rho_{i+1}(z) - \rho_i(z) \biggr]^2 \;, \label{delta1}
\end{equation}
where $\rho_i(z)$ is the density profile after the $\it i$-th
iteration, and by the quantity
\begin{equation}
\delta_2 = 1 -\frac{1}{N_s} \int^{\ell_w}_0 dz\,\rho(z) \;,
\label{delta2}
\end{equation}
accounting for the deviation from the required $N_s$.

In practice, for the calculations it is convenient to use
dimensionless variables: $z^*=z/\sigma_{ff}$ for the distance,
$\rho^*=\rho\,\sigma^3_{ff}$ for the densities, and $T^*=k_B\,
T/\varepsilon_{ff}$ for the temperature. In these units the average
density becomes $\rho^*_{av} = N_s\sigma^2_{ff}/\ell^*_w =
N^*_s/\ell^*_w$. For the numerical task, the region of integration
[0, $\ell^*_w$] was divided into a grid of equal intervals
$\Delta z^*=0.02$, i.e., a grid with 50 points per atomic diameter
$\sigma_{ff}$. It is worthwhile to notice that in the work of
Berim and Ruckenstein the number of grid points was taken equal
to 10 per atomic diameters. If the obtained profile did not change
with increasing precision from $\delta_1 \approx 10^{-8}$ to
$\delta_1 \approx 10^{-15}$, then it was accepted as a solution of
the coupled integral equations.

In a first step, we studied the same systems treated in detail by
Berim and Ruckenstein \cite{berim07}. Hence, we set $\nu=1$ and
solved the EL equation for a slit with an effective width $\ell^*_w
= 15$ at $T=87$ K ($T^*=0.73$) for a series of average fluid density
$\rho^*_{av}=N^*_s/\ell^*_w$. The ground state solutions yield
symmetric density profiles for $\rho^*_{av} < 0.1 = \rho^*_{sb1}$
and for $\rho^*_{av} > 0.514 = \rho^*_{sb2}$, while in the range
$\rho^*_{sb1} \le \rho^*_{av} \le \rho^*_{sb2}$ the ground-state
solutions provide asymmetric density profiles. This is due to the
fact that in such a regime the asymmetric solutions have lower free
energy than the symmetric ones. The free energies calculated for some
selected $\rho^*_{av}$ are listed in Table \ref{table0} together with
the results obtained by Berim and Ruckenstein \cite{berim07}. A
glance at this table indicates a good agreement between both sets
of values. In order to get symmetric solutions in the range $0.1 \le
\rho^*_{av} \le 0.513$ one must explicitly impose such a condition
to Eqs.\ (\ref{rho})-(\ref{lambda}).

\begin{figure}
\centering\includegraphics[width=8cm, angle=-0]{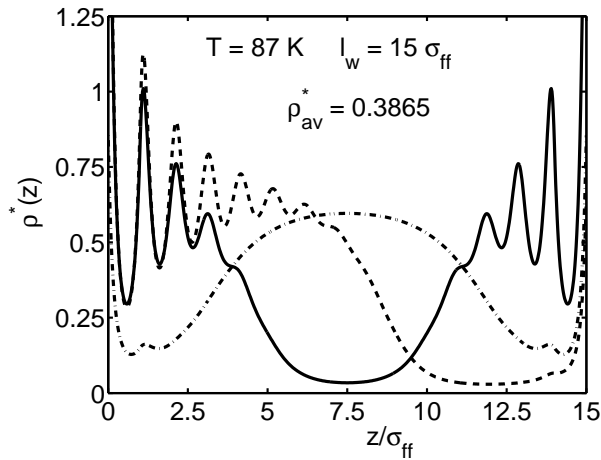}
\caption{\label{fig:rho_3865} Solutions of the EL equation
(\ref{euler1}) for the average density $\rho^*_{av}=0.3865$. The
solid curve corresponds to the lowest symmetric solution, the dashed
to the asymmetric one, and the dash-dotted to the drying-CC like
one.} 
\end{figure}

\begin{figure}
\centering\includegraphics[width=8cm, angle=0]{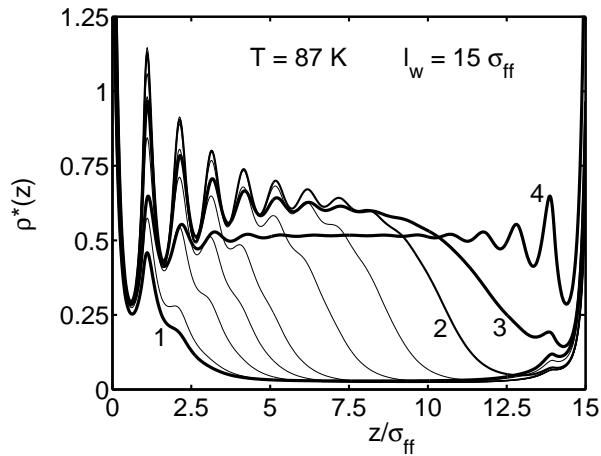}
\caption{\label{fig:profiles} Asymmetric density profiles as a
function of the distance from the hard-wall repulsion. It is shown
the evolution for increasing density average at $\nu=1$. Data for
$\rho^*_{av} = 0.1063$, 0.1159, 0.1546, 0.1932, 0.2319, 0.3092,
0.3892, 0.4638, 0.5135, and 0.5139 are displayed. These values
cover the whole range $\rho^*_{sb1} \le \rho^*_{av} \le \rho^*_{sb2}$
where there are asymmetric solutions.}
\end{figure}

\begin{figure}
\centering\includegraphics[width=8cm, height=6cm]{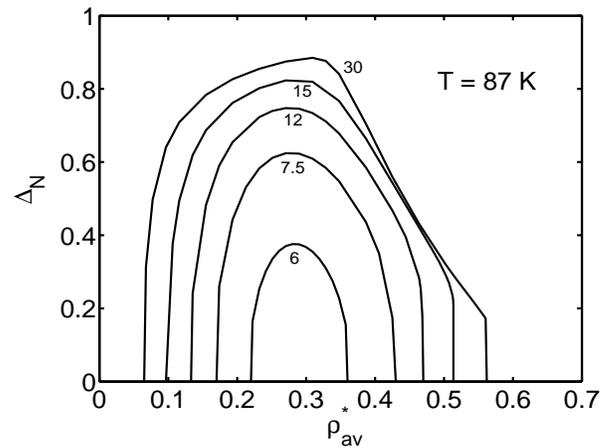}
\caption{\label{fig:L_width} Asymmetry parameter defined in Eq.\ 
(\ref{asymm}) as a function of average density. The successive
curves are results obtained for slits with different effective
widths labeled by $\ell^*_w$. For each width the asymmetric
solutions occur in different ranges $\rho^*_{sb1} \le \rho^*_{av}
\le \rho^*_{sb2}$.}
\end{figure}

The Lagrange multiplier $\mu$ (equivalent to the chemical potential
in the case of open slits) is displayed in Fig.\ \ref{fig:chemical}
as a function of average density. We show the results for a wider
range of $\rho^*_{av}$ than it is done in Fig.\ 9 of Ref.\
\onlinecite{berim07}. Figure \ref{fig:chemical} clearly indicates
that the asymmetric solutions occur in the domain where $\mu$ is
degenerate, namely, the same value of $\mu$ corresponds to different
$\rho^*_{av}$. That is just the regime where in the case of an open
slit the equal-area Maxwell construction should be applied in order
to determine the chemical potential (cf., e.g., Fig. 2 in Ref.\
\onlinecite{ancilotto99}). In addition, it is worthwhile to notice
that at $\rho^*_{sb2}$ there is an abrupt jump of $\mu$, we shall
come back to this feature below.

If one manages conveniently the EL equation it is possible to get
another kind of symmetric solutions in some region of $\rho^*_{av}$.
The free energy of such new solutions is included in Table\
\ref{table0} and the corresponding multiplier $\mu$ is plotted in
Fig.\ \ref{fig:chemical}. Figure \ref{fig:rho_3865} shows the
spacial distribution of all three states listed in Table\
\ref{table0} for $\rho^*_{av}=0.3865$. A direct comparison
indicates that these three profiles correspond to the $\it{drying}$,
the $\it{one}$-$\it{wall}$-$\it{wetting}$, and the
$\it{two}$-$\it{walls}$-$\it{wetting}$ cases displayed, respectively,
in Figs.\ 1-3 of Ref.\ \cite{nijmeijer90}. From a glance at Table\
\ref{table0} one realizes that as long as the film $\it{wetting}$
solutions exist the $\it{drying}$ one is a symmetric excited state.
In the regime $\rho^*_{av} > \rho^*_{sb2}$ the $\it{drying}$ profile
shown in Fig.\ \ref{fig:rho_3865} becomes the capillary condensation
(CC) solution.

Figure \ref{fig:profiles} shows a series of asymmetric density
profiles $\rho^*(z)=\rho(z)\,\sigma^3_{ff}$. This sequence indicates
that for increasing  average density, starting from the profile
denoted as 1 ($\rho^*_{av}=0.1063$), the number of oscillations
near the left wall as well as its amplitudes increase. This trend
continues up to the profile 2 ($\rho^*_{av}=0.4638$), where the
peaks of the oscillating structure begin to decrease. Furthermore,
profile 3 corresponds to the biggest asymmetric solution, for larger
$\rho^*_{av}$ there is a jump to the symmetric profile 4. The
latter behavior indicates a transition to the so-called CC phase,
i.e., a transition to a situation where the slit is full of liquid
argon. The jump of $\mu$ addressed in the previous paragraph is
also a manifestation of this thick-film to CC transition.

The asymmetry of the profiles for $\nu=1$ displayed in Fig.\
\ref{fig:profiles} was measured by the parameter $\Delta_N$
introduced in Eq.\ (\ref{asymm}). The results are shown in Fig.\
\ref{fig:L_width}. These values are essentially the same as that of
the equivalent curve displayed in Fig. 5 of Ref.\
\onlinecite{berim07}. As mentioned in the introduction, in Refs.\
\cite{sikkenk87,sikkenk88,nijmeijer89,nijmeijer90,velasco89,%
henderson91} it is emphasized that a slit of width $\ell^*_w=30$ is
appropriate to study wetting because it is large enough to avoid
any confinement effect. Therefore, it becomes of interest to solve
the EL equation for such a big slit and to compare the results with
that obtained for $\ell^*_w=15$. The asymmetry parameter evaluated
for the $\ell^*_w=30$ slit at $T=87$ K is also plotted in Fig.\
\ref{fig:L_width}. A comparison of the curves labeled by $30$ and
$15$ indicates that its shapes do not differ significantly, in
particular, in the regime of large coverage $0.35 \le \rho^*_{av}
\le 0.51$ the values of $\Delta_N$ are almost equal. Hence, we can
state that the capillary effects in the case of the moderately
thick slit of $\ell^*_w=15$ are not very important.

For the sake of completeness we also analyzed slits of $\ell^*_w<15$
at $T=87$ K. The asymmetry parameters evaluated for $\ell^*_w=12$,
$7.5$ and $6$ are included in Fig.\ \ref{fig:L_width}. It is clear
that these curves exhibit everywhere an increasing departure from
the $\ell^*_w=30$ results. Finally, for $\ell^*_w=5.5$ the parameter
$\Delta_N$ becomes zero for all $\rho^*_{av}$ indicating that the
asymmetric solution disappears.

\begin{figure}
\centering\includegraphics[width=8cm, height=6cm]{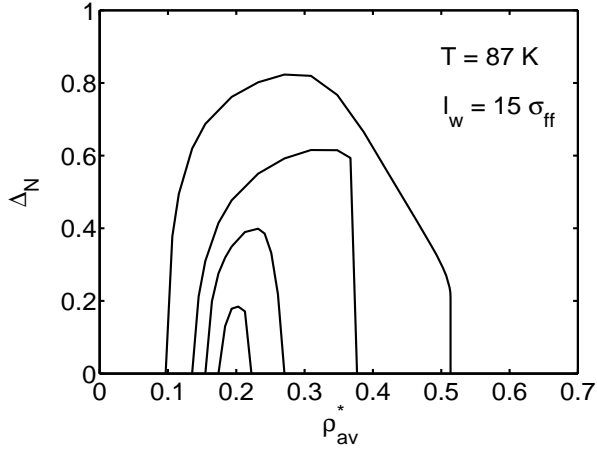}
\caption{\label{fig:asymmetry} Asymmetry parameter of density
profiles, defined in Eq.\ (\ref{asymm}), as a function of average
density. Going from the outer curve towards the inner one the data
correspond, respectively, to $\nu=1$, 1.136, 1.156 and 1.170. If
$\nu > 1.18$ for all $\rho^*_{av}$ one gets $\Delta_N=0$ .}
\end{figure}

\begin{figure}
\centering\includegraphics[width=8cm, height=6cm]{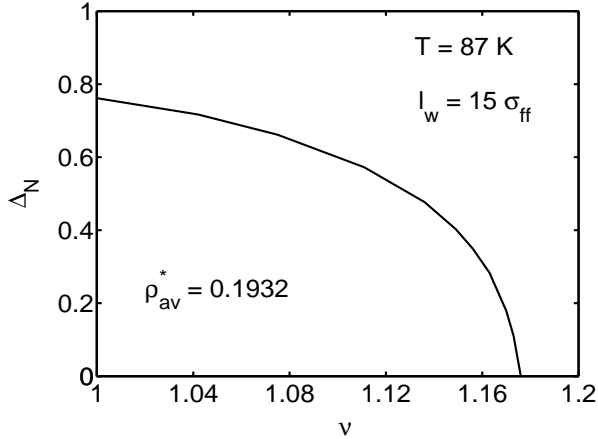}
\caption{\label{fig:asymm_1932} Asymmetry parameter of density
profiles, defined in Eq.\ (\ref{asymm}), obtained for the average
density $\rho^*_{av}=0.1932$ as a function of the factor $\nu$.}
\end{figure}

\begin{figure}
\centering\includegraphics[width=8cm, angle=-0]{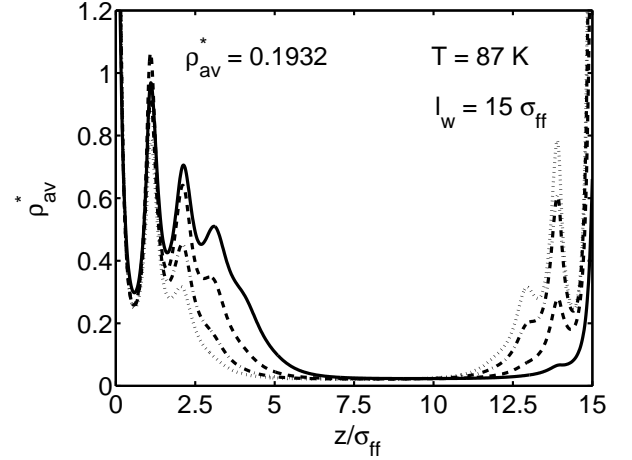}
\caption{\label{fig:rho_asymm_1932} Evolution of the asymmetric
solution for the average density $\rho^*_{av}=0.1932$ in terms of
the factor $\nu$. The solid curve corresponds to $\nu=1$, the
dashed to 1.136, the dash-dotted to 1.156, and the dotted to 1.2.} 
\end{figure}

In the next step, we analyzed the change of the properties described
above when the parameter $\nu$ is taken larger than unity. The
variation of $\nu$ is performed in such a way that the effective
width of the slit is always kept at $\ell^*_w=15$, hence, only the
potential is slightly changed mainly near the walls. At this point
the temperature was still kept at $T=87$ K. It was found that for
increasing values of $\nu$ the range of average density $\rho^*_{sb1}
\le \rho^*_{av} \le \rho^*_{sb2}$ where there are symmetry breaking
decreases. This effect might be observed in successive plots of
$\mu\,{\rm vs.}\, \rho^*_{av}$, however, we prefer to report
directly the evolution of the parameter $\Delta_N$. Figure
\ref{fig:asymmetry} shows how this parameter decreases with
increasing $\nu$. From this figure one may conclude that symmetry
breaking persists at most for $\rho^*_{av} \approx 0.2$.

Figure \ref{fig:asymm_1932} shows the asymmetry parameter
$\Delta_N$ as a function of the factor $\nu$ for the average
density $\rho^*_{av}=0.1932$. These data indicate that the
asymmetric solution already disappears for a critical value $\nu_c
\simeq 1.18$. The evolution of the density profiles from
asymmetric to symmetric species is displayed if Fig.\
\ref{fig:rho_asymm_1932}. In this drawing one may observe the
diffusion of fluid argon from the neighborhood of the left wall
towards the right one. This process continues until a symmetric
density profile is formed for $\nu \simeq 1.2$. The described
evolution of $\rho^*(z)$ is determined by the behavior of the
surface tensions. If one uses the Young's law given in Eq.\
(\ref{young}) the total surface excess energy of asymmetric profiles
may be written as
\begin{equation}
\gamma_{tot} = \gamma_{SL} + \gamma_{LV} + \gamma_{SV} 
= 2\,\gamma_{SL} + \gamma_{LV}\,(1 + \cos{\theta}) \;,
\label{phase}
\end{equation}
with $\cos{\theta}=(\gamma_{SV}-\gamma_{SL})/\gamma_{LV}<1$. By
increasing enough the attraction the equality $\gamma_{SV}-
\gamma_{SL}=\gamma_{LV}$ is reached yielding $\cos{\theta}=1$, then
the system undergoes to a transition to a symmetric profile with
\begin{equation}
\gamma_{tot} = 2\,\gamma_{SL} + 2\,\gamma_{LV} \;.
\label{phases}
\end{equation}
In this case both walls of the slit are wet.

Let us now look if something special is going on for the
adsorption potential at $\nu_c \simeq 1.18$. As a matter of fact,
it becomes important to explore where the hard-wall repulsion
introduced in Ref.\ \onlinecite{berim07} is located for such a
critical value. The position of the left hard-wall repulsion
with respect to the real wall according to Fig.\
\ref{fig:potential} is given by
\begin{equation}
\xi^c_{\rm HW} = \frac{\sigma_{sf}}{\nu_c} \;. \label{ksic}
\end{equation}
For a slit of effective width $\ell^*_w=15$ the minimum of the
total adsorption potential given by Eq.\ (\ref{snilson}) is to a
very good approximation determined by the minimum of the ES
potential exerted by the left wall. It is located at
\begin{equation}
\xi_{min} = \left(\frac{2}{5}\right)^{1/6}\,\sigma_{sf} \;.
\label{ksim}
\end{equation}
The ratio of these quantities becomes
\begin{equation}
\frac{\xi_{min}}{\xi^c_{\rm HW}} = \left(\frac{2}{5}\right)^{1/6}
\,\nu_c \simeq 1.013 \;. \label{ksir}
\end{equation}
Since this ratio is larger than unity, it indicates that for
$\nu_c$ the minimum of the ES potential is reached inside the slit.

We also analyzed the occurrence of asymmetric solutions for $1
\le \nu \ < 2$ at temperatures $T > 87$ K. It was also found that
the symmetry breaking disappears at some critical value of $\nu$.
Such a behavior may be expected if one takes into account that
for temperatures larger than $T=87$ K even for $\nu=1$ the range
$\rho_{sb1} \le \rho_{av} \le \rho_{sb2}$ is smaller (see Fig. 5
of Ref.\ \onlinecite{berim07}).

It is worthwhile to notice that a reliable {\em ab-initio} potential
utilized by Mistura {\it et al.} \cite{mistura99} for investigating
the adsorption of Ar on CO$_2$ exhibits an even stronger attraction
than the ES potential. This feature can be observed in Fig.\ 3 of
\cite{mistura99}. Hence, one would not expect any symmetry breaking
in the case of such a realistic adsorption potential.

\section{Final remarks}
\label{sec:conclude}

We reexamined the symmetry breaking found very recently by Berim
and Ruckenstein \cite{berim07} in a study of the adsorption of
argon in a closed slit with identical walls of carbon dioxide.
It is important to stress that these authors have introduced
hard-wall repulsions at distances $d_{BR}=\sigma_{sf}$ from the
real walls of the slit as shown in Fig.\ \ref{fig:potential} (see
also Fig. 2 in Ref.\ \cite{berim07}), reducing in such a way the
strength of the adopted ES potential. Stable asymmetric solutions
were obtained in the case of a slit with effective width $\ell^*_w
=15$ for temperatures in the range $87 \le T \le 106$~K.

The present study was mainly devoted to establish how robust are
the asymmetric solutions against changes of the adsorption potential.
However, in addition, the behavior of the asymmetry parameter was
also evaluated for slits of different widths. The calculations have
been carried out using the same nonlocal formalism as that adopted
in Ref.\ \onlinecite{berim07}, namely, the SDA density functional
theory proposed by Tarazona \cite{taraz85}.

Since in a pioneering series of works a Dutch Collaboration have
previously found the symmetry breaking by studying adsorption in a
slit of width $\ell^*_w=30$, see Refs.\ \cite{sikkenk87,%
sikkenk88,nijmeijer89,nijmeijer90}, we performed a comparison of
the asymmetry parameter $\Delta_N$ evaluated for slits of $\ell^*_w=
15$ and $30$ at $T=87$ K. The difference between the results is
rather small. Further calculations showed that the effect due to
confinement begins to be important for slits with $\ell^*_w \le 12$.

Focusing the analysis on the slit of $\ell^*_w=15$, in a rather
complete plot of $\mu\,{\rm vs.}\,\rho^*_{av}$ we show clearly the
regime where the symmetry breaking occurs and, in addition, we also
display the Lagrange multiplier $\mu$ for a CC like solution.
Furthermore, it is shown that the spacial shape of the obtained
solutions correspond to the three sorts of profiles discussed in
Refs.\ \cite{sikkenk87,sikkenk88,nijmeijer89,nijmeijer90}. The
strength of the attraction determines which one of that profiles is
the stable solution.

By shifting the hard-wall repulsions introduced by Berim and
Ruckenstein \cite{berim07} towards the real walls the strength of 
the attraction is increased producing changes in the balance of
surface tensions. It was found that at $T=87$~K already for the
critical distance $d_c \simeq d_{BR}/1.2$ the asymmetric solutions
disappear. At this temperature, close to the triple point, the
asymmetry parameter $\Delta_N$ reached its largest values in the
study of Ref.\ \onlinecite{berim07}. For higher temperatures the
symmetry breaking disappears even more rapidly.

Finally we can state that from the present study it is possible to
conclude that the symmetry breaking reported in Ref.\
\onlinecite{berim07} can be understood in terms of findings described
in Refs.\ \cite{sikkenk87,sikkenk88,nijmeijer89,nijmeijer90}. By
locating hard-wall repulsions the authors of Ref.\
\onlinecite{berim07} diminish the attraction of the ES adsorption
potential exerted on the fluid causing the entrance of the system
in a $\it{one}$-$\it{wall}$-$\it{wetting}$ regime (asymmetric
profiles) corresponding to Fig.\ 2 of Ref.\ \cite{nijmeijer90}. By
moving the hard-wall repulsions towards the real walls the attraction
increases monotonically causing eventually the entrance of the
system in the $\it{two}$-$\it{walls}$-$\it{wetting}$ regime with
symmetric solution of the type displayed in Fig.\ 3 of Ref.\
\cite{nijmeijer90}.

Furthermore, a reliable {\it ab-initio} potential for the
interaction of fluid Ar with a structureless smooth wall of CO$_2$
like that adopted by Mistura {\it et al.} \cite{mistura99} is even
stronger than the ES one. Therefore, it is possible to infer that
no symmetry breaking would be expeted for real Ar/CO$_2$ systems.

\begin{acknowledgments}
This work was supported in part by the Ministry of Culture and
Education of Argentina through Grants CONICET PIP No. 5138/05, 
ANPCyT BID 1728/OC - PICT No. 31980, and UBACYT No. X298.
\end{acknowledgments}

\end{document}